\documentclass[twocolumn,twoside,dvips,showpacs,preprintnumbers,aps,prl,amsmath,amssymb,longbibliography]{revtex4-1}

\usepackage{graphicx}
\usepackage{grffile}

\usepackage{dcolumn}
\usepackage{bm}
\usepackage{amsmath}
\usepackage{amssymb}
\usepackage{bbm}
\usepackage{mathrsfs}
\usepackage[sans]{dsfont}
\usepackage{epsfig,psfrag}
\usepackage{amsmath,amsfonts,amssymb}
\usepackage[usenames]{color}
\usepackage[dvips, bookmarks, colorlinks=true, plainpages = false, citecolor = blue, linkcolor = blue, urlcolor = blue, filecolor = blue]{hyperref}

\newcommand{\gd}{\ensuremath{\dot\gamma}}
\newcommand{\ie}{{ i.e.,~}}

\newcommand{\sr}{\ensuremath{\dot\gamma}}
\newcommand{\sxy}{\ensuremath{\sigma_{xy}}}

\newcommand{\beq}{\begin{equation}}
\newcommand{\eeq}{\end{equation}}

\newcommand{\rmd}{\mathrm{d}}

\begin{document}



\title{Characterizing rare fluctuations in soft particulate flows}

\author{S. H. E. Rahbari$^1$} \email{habib.rahbari@gmail.com}
\author{A. A. Saberi$^{2,3,4}$}\email{ab.saberi@ut.ac.ir}
\author{Hyunggyu Park$^1$} 
\author{J. Vollmer$^{5,6,7}$}

\affiliation{$^1$ School of Physics, Korea Institute for Advanced Study, Seoul 130-722, South Korea} 
\affiliation{$^2$ Department of Physics, College of Science, University of Tehran, P.O. Box 14395-547, Tehran, Iran}
\affiliation{$^3$ School of Physics and Accelerators, Institute for research in Fundamental Science (IPM), P.O. 19395-5531, Tehran, Iran }
\affiliation{$^4$ Institut f\"ur Theoretische Physik, Universitat zu K\"oln, Z\"ulpicher Strasse 77, 50937 K\"oln, Germany}
\affiliation{$^5$ Max Planck Institute for Dynamics and Self-Organization (MPI DS), 37077 G\"ottingen, Germany}
\affiliation{$^6$ Faculty of Physics, Georg August University G\"ottingen, 37077 G\"ottingen, Germany}
\affiliation{$^7$ Faculty of Mathematics and Computer Sciences, Georg August University G\"ottingen, 37073 G\"ottingen, Germany}

\begin{abstract}

  {

    Soft particulate media include a wide range of systems involving
    athermal dissipative particles both in non-living and biological
    materials. Characterization of flows of particulate media is of
    great practical and theoretical importance. A fascinating
    feature of these systems is the existence of a critical rigidity
    transition in the dense regime dominated by highly intermittent
    fluctuations that severely affects the flow properties. Here, we
    unveil the underlying mechanisms of rare fluctuations in soft
    particulate flows. We find that rare fluctuations have different
    origins above and below the critical jamming density and become
    suppressed near the jamming transition. We then conjecture a
    time-independent local fluctuation relation, which we verify
    numerically, and that gives rise to an effective temperature. We
    discuss similarities and differences between our proposed
    effective temperature with the conventional kinetic temperature in
    the system by means of a universal scaling collapse.


  }

\end{abstract}

\maketitle


\section{Introduction}

Large fluctuations are a distinguishing feature of soft particulate
flows, like flows of granular media \cite{miller_1996, peng_1995},
bubbles and foams \cite{durian_1995}, and in living matter such as
biological tissues \cite{angelini_2011, bi_2015}.  Very dense systems
are in a jammed state.  They only move in response to a strong
external force. Less packed systems are in a fluid state.  They flow
in response to any finite force.  The flows are highly intermittent
and involve rare, very large fluctuations \cite{tighe_2010} that can
trigger transitions between the jammed and the fluid state
\cite{gennes_1999}.  Landslides \cite{jaeger_1996} and avalanches
\cite{schweizer_2003} are transitions from a jammed to a fluid state.
Clogging of hoppers \cite{zuriguel_2014} and breakdown of silos
\cite{dogangun_2009} involve the transition from a fluid to a jammed
state.  Predicting the frequency of appearance of such fluctuations is
a question of great practical and theoretical interest.

Fluctuation relations (FRs) compare the probability of the forward
progression of a dynamics and its reverse; akin of watching a movie
played in forward and reverse direction.  They provide an exact
symmetry property of the probability distribution function
characterizing the likelihood to encounter a given course of states in
an observation of the dynamics.  Close to equilibrium this symmetry
entails linear response.  Far from equilibrium FRs have been adopted
in micro-biological systems to determine the free energy of a folding
RNA \cite{collin_2005} and thermodynamic properties of other
biomolecules \cite{seifert_2012, parrondo_2015}. In contrast to the
dynamics of the microscopic biological systems the dynamics of most
macroscopic systems do not move against an exerted
force~\cite{toyabe_2010}. The strongly fluctuating and intermittent
flows of soft particulate matter are a noticeable exception to this
rule.

Here, we analyze the statistics of those very large fluctuations where
the flow is moving against a driving force. We discuss rare
fluctuations in flows of soft particulate matter, where the injected
power, $p=\rmd w/\rmd t\propto \sxy \cdot \delta v$ takes negative
values in a finite domain that is subjected to a velocity gradient
$\delta v$ and that resists flow by a shear stress $\sxy$ (the shear
stress is the force resisting the flow, see Supplementary Notes 1, 2,
and 3). In a steady state the injected energy balances the energy
dissipated by the viscosity of the fluid.  Hence, on average $p$ takes
a positive value, and in the thermodynamic limit it does not
fluctuate.  When there is a finite number of particles in the
considered domain there is a small chance to encounter {rare
  fluctuations} where $p$ takes a negative value.  This can either be
due to the reversal of the shear stress $\sxy$ or to the velocity
gradient $\delta v$.  While one might naively expect that fluctuations
in $\sxy$ and $\delta v$ would equally contribute to such violations,
our numerical simulations show an unexpected interplay of these two
mechanisms of rare fluctuations. Moreover, we establish a {
  variation of} fluctuation relation (FR) for the statistics of the
injected power driving the flow and use it to define an effective
temperature for far-from-equilibrium soft particulate flows.{ 
  Our approach can be easily generalized to study negative power
  fluctuations and effective temperatures both in simulations and
  experiments in a wide range of problems such as in the sheared
  foams, vibrated granular media, particles down an inclined plane,
  emulsions and other soft particulate media.

}

\section{Results}

\begin{figure}[h]
  \centering
  \hfill\includegraphics[width=0.475\textwidth]{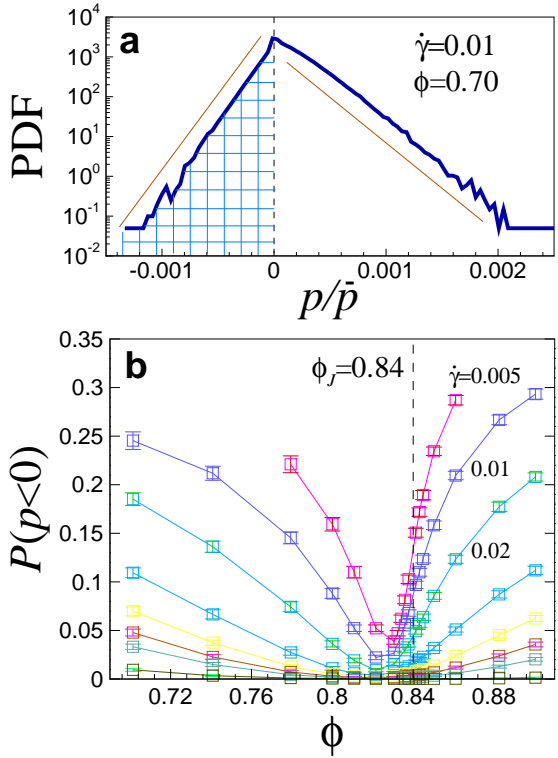}
  \hfill\\
  \caption[]{\textbf{Probability of rare fluctuations.}  ({\bf a}) A
    typical probability distribution function (PDF) of the rescaled
    power $p/\bar{p}$ for $\phi=0.7$, $\dot{\gamma}=0.01$, and
    $L=30$. The solid straight lines show the exponential decay of the
    PDF for large and small arguments. The area of the shaded region
    gives the probability, $P(p<0)$, that the local power takes a
    negative value, i.e., to encounter a negative power
    injection. ({\bf b}) The probability to observe a negative power
    injection $P(p<0)$ as a function of packing fraction $\phi$ for
    different shear rates $\dot{\gamma} = 0.005, 0.01, 0.02, 0.04,
    0.06, 0.08, 0.1$ and $0.2$ from top to bottom, respectively, and
    system size $L=30$. The vertical dashed line marks the critical
    packing fraction, $\phi_J$, where the jamming transition occurs in
    the static limit. { Error bars correspond to square root
      of variance.}
    \label{fig:prob_total}}
\end{figure}

{\bf Probability distribution function of injected power $p$.} In the
Fig.~\ref{fig:prob_total}-a, we show a typical example of the
probability distribution function (PDF), $\mathcal{P}(p/\bar{p})$, of
the local power flux rescaled by the mean power, i.e., $p/\bar{p}$.
The PDF exhibits several remarkable features. The power flux can take
negative values with a rather high probability. The distribution is
strongly skewed towards positive events.  At the both sides the PDF
decays exponentially to a good approximations.  It is very different
from a Gaussian distribution.  Still, the negative part of the PDF
(the shaded area in  Fig.~\ref{fig:prob_total}-a) decreases
rapidly with system size.  In the following this area will be denoted
as $P(p<0) \equiv \int_{-\infty}^0 \mathcal{P}(p) \: \rmd p$. The
slopes of the exponential decay are roughly proportional to the number
of particles in the considered volume such that $P(p<0)$ decays
exponentially to zero with system size.


{\bf Probability of rare fluctuations.}  In
Fig.~\ref{fig:prob_total}-b we show $P(p<0)$ as a function of packing
fraction $\phi$.  The lines in different color indicate data for
different shear rates, $\sr$.  For shear rates, $\sr \gtrsim 0.2$
(olive), this probability is very small, and it grows upon decreasing
the shear rate.  There is pronounced growth in the fluid and in the
jammed states.  However, close to the jamming point $\phi_J$ (marked
by the vertical dashed line) it remains small for all shear rates.
The critical point lies to the right of the minima.  However, the
minima converge towards $\phi_J$ in the limit $\dot{\gamma}\rightarrow
0$ and system size $L \to \infty$.  Hence, the minimum will eventually
approach $\phi_J$; the discrepancy is due to finite-size effects.  The
emergence of the global minimum is remarkable because fluctuations are
expected to diverge close to a critical point \cite{olsson_2007,
  hatano_2008, hayakawa_2013}, i.e., one expects a larger probability
to encounter fluctuations close to the critical point.  The minimum is
a distinctive feature of the jamming transition.  It has no
counterpart in equilibrium thermodynamics.

\begin{figure}[b]
  \centering
  \hfill\includegraphics[width=0.5\textwidth]{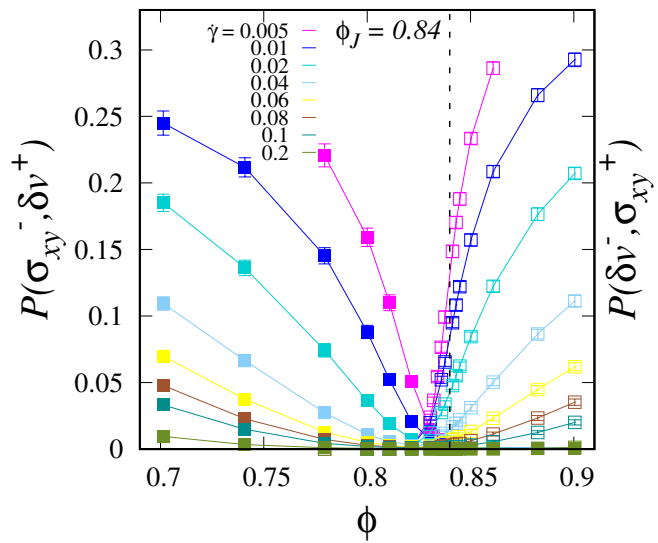}
  \hfill\\
  \caption[]{{\bf Mutually exclusive fluctuations.}  Joint
    probabilities
    $P(\sigma_{xy}^{-}, \delta v^{+})$ (left axis, filled symbols) and
    $P(\delta v^{-}, \sigma_{xy}^{+})$ (right axis, hollow symbols)
    as functions of packing fraction, $\phi$, for
    various shear rates, $\sr$.  In the fluid state, $\phi<\phi_J$,
    the dominant mechanism of negative power injection is the
    reversal of the shear stress.
    In the jammed state, $\phi>\phi_J$, it is due to the reversion of the velocity
    gradient. { Error bars correspond to
      square root of variance.}
    \label{fig:prob_joint}}
\end{figure}

{\bf Decomposition of the probability of rare fluctuations.}  In order
to understand the origins of the negative power injection, we recall
that the power flux has two contributions: The local shear stress
$\sigma_{xy}$, and the local velocity gradient $\delta v$.  Negative
power injection arises whenever
either $\sigma_{xy}<0$ and $\delta v > 0$,
or $\delta v < 0$ and $\sigma_{xy} > 0$. 
The former events will be denoted as $(\sigma_{xy}^{-}, \delta v^{+})$.
In this case the velocity profile remains monotonic and negative power injection arises from fluctuations of the shear stress.
The latter events will be denoted as $(\delta v^{-}, \sigma_{xy}^{+})$.
In this case the negative power injection is connected to rare fluctuations where the velocity profile is no longer monotonic ( see Supplementary Fig.~1).
The joint probability of these  events sum up to the probability to encounter negative power injection:
$P(p<0)
=P(\sigma_{xy}^{-}, \delta v^{+})
+P(\delta v^{-}, \sigma_{xy}^{+})$ (In Supplementary Note 5 and Supplementary Fig.~3 we numerically prove this equality).
Figure~\ref{fig:prob_joint} presents the numerical results for
these two joint probabilities, i.e., $P(\sigma_{xy}^{-}, \delta
v^{+})$ (left axis, filled symbols) and $P(\delta v^{-},
\sigma_{xy}^{+})$ (right axis, hollow symbols) as function of packing
fraction $\phi$ for various shear rates.  For a given shear rate (a
given color), the intersection point of the joint probabilities is
very close to the jamming point.  Hence, $\phi_J$ splits the
probability space into two disjoint regions.  Accordingly, the two
types of mechanisms, shear-stress and velocity-gradient fluctuations,
are {mutually exclusive}. The reason can be sketched as follows. The
positive shear stress corresponds to head-to-head collisions while the
negative shear stress is associated to backup collisions during which
the local angular momentum is in the same and in the opposite
direction of the global induced angular momentum of the flow,
respectively (see Supplementary Fig.~2 and Supplementary Note~4 ).
For $\phi<\phi_J$, since the average coordination number is relatively
small, the negative contacts can be observed with high probability
although they are suppressed by increasing the shearing flow. This
justifies the decreasing dependence of the probability of negative
stress as function of both shear rate and packing fraction.  For
$\phi>\phi_J$, since the coordination number jumps to a value $\ge 4$,
the negative contacts can in average be suppressed by the positive
ones since the global symmetries of the flow favor the positive
contacts. This elucidates the behavior seen in
Fig.~\ref{fig:prob_joint} for the negative shear stress and since the
negative power includes exclusive contributions from the negative
shear stress and negative velocity gradient, our argument naturally
explains the observation of the mutually exclusive fluctuations.



{\bf Fluctuation relation.}
At this point we identified qualitatively different physics underlying the fluctuations of fluid and jammed systems.
In order to gain more insight into the parameter dependence of the strength of fluctuations we establish now a fluctuation relation for the flows.
It will characterize the width of the PDFs $\mathcal{P}(p)$ by an effective temperature $T_e$.
We conjecture that the relation
$\ln[P(p)/P(-p)]=\beta p$ holds where $\beta=\tau/T_e$ has inverse
dimension of power.
Here, the constant $\tau$ is the relevant elastic time scale which
represents the typical time scale of a single collision.  It is
approximately independent of $\phi$ and $\sr$ (see Supplementary
Note~6).


Figure~\ref{fig:FT} presents the numerical verification of our
conjecture for two packing fractions $\phi=0.7$ and $0.9$ that lie
below and above the critical point, $\phi_J=0.84$, respectively.
There is a linear dependence between $\ln[P(p)/P(-p)]$ and $p$ whose
slope is a decreasing function of the shear rate $\dot{\gamma}$.  This
implies that $T_e$ is an increasing function of the shear rate ---in
accordance with the shear-rate dependence of the average kinetic
temperature of particles in the flow.  Surprisingly, the
correspondence is not only qualitative.  It even holds quantitatively.
All data shown in Fig.~\ref{fig:FT}-a collapse to a straight line when
$\ln[P(p)/P(-p)]$ is multiplied by the granular temperature, $T_g$
(Fig.~\ref{fig:T_eff_master}-b).  The resulting straight line has a
slope $1$ with $\tau=0.28$. ({In Supplementary Note~8 and Supplementary Fig.~8, we
  numerically prove that our proposed FR is also satisfied for highly
  damped systems corresponding to non-Brownian suspensions.})



{\bf Scaling collapse of effective and kinetic temperatures.}  In the
jammed state the effective temperature, $T_e$, and the granular
temperature, $T_g$, differ: $T_e$ is always larger than $T_g$.  We
explore the parameter dependence of the two temperatures by exploring
their scaling properties \cite{vagberg_2016}.  In
Fig.~\ref{fig:T_eff_master}-a we demonstrate that the full $\sr$ and
$\phi$ dependence of the temperatures { for different system sizes}
can be represented in terms of a master plot where $T /
|\delta\phi|^y$ is plotted as function of $\sr/|\delta\phi|^{y/q}$
with appropriate choice of the exponents $y$, $q$ and $\delta\phi =
\phi-\phi_c$.  For the fluid state we thus find the well-known
Bagnoldian scaling with exponent $2$.  In the jammed state, we find
that $T_e$ and $T_g$ still collapse uniformly in the critical region,
$\sr/|\delta\phi|^{y/q} \gtrsim 10$.  For jammed flows, however, they
segregate into two different branches in accord with our earlier
statement.  In this limit $T_e$ approaches a constant---yield stress
emerges.  In contrast, $T_g$ shows a power-law behavior with exponent
$\sim 1.5(1)$. { We have checked consistency of all our exponents, \ie
  exponents of granular temperature and components of stress tensor,
  in Supplementary Note~7 and Supplementary Fig.~4-7. We also show in
  Supplementary Note~8 that these exponents are universal in a sense
  that the same scaling collapse is achieved with the same critical
  exponents for highly dissipative regime ---in connection with the
  non-Brownian suspensions.}\\




{

Heussinger { et al.}~\cite{heussinger_2009, heussinger_2010} have
studied fluctuations of some observables in the flow of an assembly of
frictionless, soft discs at zero temperature, in the vicinity of and
slightly above $\phi_J$. They have found that the contact-number
fluctuations and relative fluctuations of the shear stress diverge
upon approaching $\phi_J$ from above. They also report on strong
finite-size effects when using $\phi$ as control parameter. However,
the effective temperature $T_e$ in our study, is the product of
thermodynamically forbidden fluctuations (negative stress below
$\phi_J$ and negative velocity gradient above $\phi_J$) which are
specific to small-size systems and vanish in the vicinity of
$\phi_J$. Our observation indicates that the scaling behavior of $T_e$
is not significantly altered by the finite-size effects below and
above $\phi_J$ for $L>10$ ---see Fig.~\ref{fig:T_eff_master}-a. We
find that to a very good extent, $T_e$ is independent of the system
size. We only see small deviations for $L=10$. These deviations vanish
as $\dot\gamma\rightarrow 0$.}




\section{Discussion}

The flow of particulate matter is similar to classical fluids in so
far as it involves the motion of many particles that interact by
short-range forces.
As function of packing fraction the flows undergo a phase transition
from a fluid into a jammed state.  Close to the critical point the
materials obey scaling relations \cite{paredes_2013, dinkgreve_2015},
reminiscent of critical phenomena.
The data collapse of the granular temperature, i.e., the kinetic
energy per degree of freedom, is shown here in
Fig.~\ref{fig:T_eff_master}-a.  In the fluid state and in the critical
region this temperature agrees with an effective temperature that
characterizes the probability to encounter different power injections.

The proposed effective temperature is sensitive to the inherent
properties of the systems, and it potentially qualifies as the
effective temperature that has been searched for recently with great
urgency~\cite{makse_2002, behringer_2002}.  The effective temperatures
proposed in the past \cite{ono_2002} are based on
fluctuation-dissipation relations, i.e., they assume linear-response.
Our study goes beyond linear response by introducing a fluctuation
relation in order to define a shear-rate dependent effective
temperature.  This effective temperature is valid for packing
fractions far from the jamming point, in contrast to the previous ones
that are meaningful measure only near the transition
point~\cite{ono_2002}.\\

{ Various types of fluctuation theorems have been
  extensively studied over the last two decades~\cite{ jarzynski_1997,
    crooks_1999, jarzynski_2011, seifert_2012}. Motivated by molecular
  dynamics simulations, Evans { et al.}~\cite{evans_1993} proposed an
  empirical fluctuation relation for entropy production rate in a
  $2$-dimensional sheared Lennard-Jones fluid. Later, this empirical
  relation was rigorously proved by Gallavotti and
  Cohen~\cite{gala_1994, gala_1995}. This is now known as steady-state
  fluctuation theorem (SSFT). In a steady-state fluctuation theorem,
  the entropy production rate is time averaged over a single, randomly
  sampled interval of duration $\tau$. In contrast, the transient
  fluctuation theorem (TFT) of Evans and Searles~\cite{evans_1994}
  applies to a system that evolves from an initial equilibrium state
  to a nonequilibrium steady state. TFTs are different from SSFTs from
  a practical point of view. Whereas TFRs rely on ensemble averaging
  all starting from the same initial macro-state, SSFRs may be
  verified from steady-state evolution of a system over a sufficiently
  long time~\cite{marconi_2008}. As we have already stressed out,
  properties of soft particulate flows are predominated by the rare
  fluctuations which result to intermittent behavior of these
  flows. It can be seen that fluctuation relations of type SSFTs are
  not suitable for soft particulate flows. The reason is that as a
  consequence of averaging process during the sampling time, the rare
  fluctuations can be washed out. Therefore, we use an instantaneous,
  time-independent fluctuation relation to characterize strength of
  rare fluctuations. Whether our postulated fluctuation relation would
  enjoy a rigorous treatment, will remain a theoretical challenge.}



In addition, we have shown here that fluctuations in soft particulate
flows differ essentially from those of classical fluids. First of all,
they are very strong as demonstrated by the exponential decay of the
PDFs of negative power injection (Fig.~\ref{fig:prob_total}-a) rather
than the much faster decay of Gaussian distributions.  Even for shear
rates as large as $\sr=0.2$ we observe negative power injection
(lowermost curve in Fig.~\ref{fig:prob_total}-b).
Even more surprising, rare fluctuations are strongly suppressed close
to the critical point (minima of the curves in
Fig.~\ref{fig:prob_total}-b).  They behave exactly contrary to the
strength of critical fluctuations that diverge at the critical point
and die out rapidly outside the critical region \cite{olsson_2007}.
It will be challenging numerically, but extremely interesting from a
conceptual point of view to explore how classical fluids behave in
this respect.
Finally, we have shown that there are different physical mechanisms
underlying rare fluctuations in fluid and jammed states: In fluid
states negative power injection originates from fluctuations where the
shear stress takes a negative sign ---in jammed states they arise in
regions with negative shear rates.
This dichonometry of mutually exclusive mechanisms of rare
fluctuations above and below a critical state is a distinctive feature
of soft particulate flows that has no counterpart in equilibrium
thermodynamics.

Rheological properties of particulate flows are commonly characterized
in terms of hydrodynamic equations and constitutive relations
\cite{jop_2006, sollich_1997}.  Fluctuations are not a part of the
modeling.  The present study takes a different approach to
characterize the systems: We focus on the fluctuations as an inherent
property of the dynamics.  Remarkably, the fluctuations obey a local,
time-independent fluctuation relation, and this relation can be used
to define an effective temperature of the system.  In contrast to the
hydrodynamic approaches the temperature is {not} a field variable in
this setting. Rather it characterizes the variability of snapshots of
the flow. It is a scalar quantity that characterizes the ensemble of
observations of the flows, taking full note of fluctuations.  It
neither requires to find appropriate heuristic constitutive equations,
nor does it rely on the scale separation at mesoscopic scales that is
implicit to the definition of thermodynamic fields.  Hence, it is less
prone to pitfalls arising from inapt choices of constitutive relations
and applicable to a larger class of far-from-equilibrium flows.  Thus,
the present study opens a qualitatively new road to the description of
far-from-equilibrium particulate flows.\\

{ In a recent study, Maloney~\cite{maloney_2015}
  investigated distribution of dissipation power $\mathcal{P}(\Gamma)$
  in Durian's bubble model. Whereas we find that $\mathcal{P}(p)$ has
  always exponential tails, it is shown that above jamming density
  $\mathcal{P}(\Gamma)$ becomes power law for small shear rates. This
  implies that distributions of injection and dissipation powers might
  not be equivalent.  Stationarity condition implies that first
  moments of these distributions must be equal. But as one can see,
  higher moments of these distributions, which refer to the
  characteristic of the tails, might be different.  This suggests a
  new avenue of research for investigation of steady state properties
  of non-equilibrium states.  In this context, Maloney's
  work~\cite{maloney_2015} together with our approach to calculate
  $\mathcal{P}(p)$ in shear flows provide a solid framework for
  investigation of similarities and differences between distributions
  of injection and dissipation powers.}

%
%
%
%
%
%


\section{Methods}
\label{Sec:methods}

We perform molecular-dynamics simulations of two-dimensional frictionless
bidisperse disks.
Particles interact via short range repulsive and dissipative forces.
Two particles $i$ and $j$ of radii $R_i^a$ and $R_j^b$
(where $a,b=0$, $1$ stand for two different radius of bidisperse particles)
at positions ${\bf r}_i$ and ${\bf r}_j$ interact when
$\xi_{ij} = R_i^a + R_j^b - r_{ij} > 0$.
Here
$\xi_{ij}$ is called the mutual compression of particles $i$ and $j$,
$r_{ij} = |{\bf r}_i - {\bf r}_j|$.
The particles interact via a linear Dashpot model,
$F_{ij} = Y \xi_{ij} + \gamma \frac{d\xi_{ij}}{dt}$,
where $Y$ and $\gamma$ are denoted as elastic and dissipative constant, respectively.
Throughout the study we adopt the values
$Y = 100$ and $\gamma=0.315$, respectively.

In order to prevent crystallization we use a $1:1$ binary mixture of particles 
where the ratio of the radii of large and small particles is set to $R^1/R^0=1.4$.

The equations of motion are non-dimensionalized by choosing 
the unit of the length to be $R^0 + R^1 = 1$, and
setting the mass of each particle equal to its area, $m_a = \pi [R^a]^2$.
Finally, the ratio of $Y$ and $\gamma$ provides the time scale
$t^\star = \gamma / Y = 3.15\times10^{-3}$.

Lees-Edwards boundary conditions are applied along ${ x}$-direction.
They create a uniform overall shear rate,~$\gd$.
These equations of motion are integrated with a $5^\textrm{th}$-order
predictor-corrector Gear algorithm with time step, $dt = 10^{-4}$.

{\bf Acknowledgments.}  Fruitful discussions with J. Nagler and
M. Schr\"oter are highly acknowledged. S. H. E. R. thanks the Korea
Institute for Advanced Study for providing computing resources (KIAS
Center for Advanced Computation - Linux cluster system) for this work,
and specially consultations from Hoyoung Kim. A.A.S. would like to
acknowledge supports from the Alexander von Humboldt Foundation, and
partial financial supports by the research council of the University
of Tehran. {We also would like to thank referees of
  Nat. Comm. for fruitful comments which eventually increased quality
  of our manuscript.}\\

{
{\bf Data availability.} The authors declare that the data supporting
the findings of this study are available from the authors on request.
}

\section{Contributions}

S.H.E.R. performed the simulations.  All authors discussed the data,
participated in the data analysis, and in writing the manuscript.  The
first two authors had equal contributions in the development of the
study.  \\

\section{Competing interests}
The authors declare no competing financial interests.



\begin{figure}
  \centering
  \hfill\includegraphics[width=0.5\textwidth]{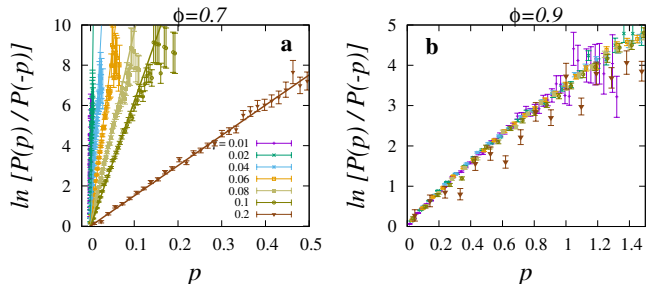}
  \hfill\\
  \caption[]{\textbf{Verification of the instantaneous fluctuation
      relation.} Plot of $\ln[P(p)/P(-p)]$ vs $p$ for two packing
    fractions ({\bf a}) $\phi=0.7$ and, ({\bf b}) $\phi=0.9$. The
    solid lines are linear fits of slope $\beta_e=\tau/T_e$ of the
    data for different shear rates. The slope decreases by increasing
    the shear rate $\dot{\gamma}$, implying that the effective
    temperature $T_e$ increases by $\dot{\gamma}$. The slope has a
    weak dependence on $\dot{\gamma}$ in the jammed state. {
      For $n^+$ and $n^-$ representing number of positive $(+p)$ and
      negative $(-p)$ cases, the corresponding error bar of
      $P(p)/P(-p)$ is equal to $(1/n^+ + 1/n^-)^{1/2}$.}
    \label{fig:FT}}
\end{figure}

\begin{figure}
  \centering
  \hfill\includegraphics[width=0.45\textwidth]{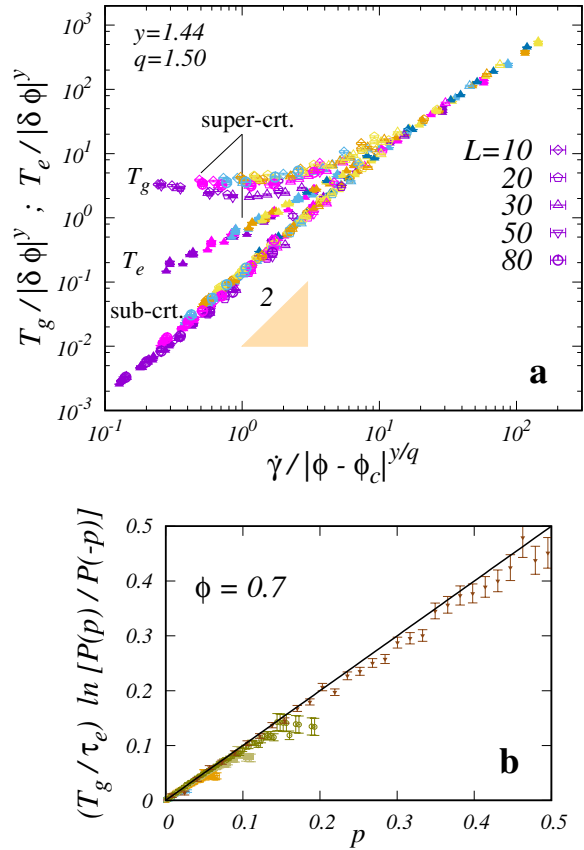}
  \hfill\\
  \caption[]{\textbf{Scaling of effective and kinetic temperatures.}
    ({\bf a}) When rescaled with the critical exponents $q=1.5(1)$ and
    $y=1.44(15)$ the effective and granular temperatures collapse onto
    a scaling function.  However, to achieve a data collapse we had to
    adopt slightly different critical densities, $\phi_c=0.83$ and
    $\phi_c=\phi_J=0.84$ for $T_e$ and $T_g$, respectively.  In the
    fluid state and in the critical state the temperatures match.  For
    the fluid state they exhibit Bagnoldian scaling with exponent $2$.
    In the critical state they still share same non-trivial scaling
    for $\dot{\gamma}/\delta\phi^{y/q} \gtrsim 10$.  In the jammed
    state the temperatures segregate into two different branches;
    $T_e$ approaches a constant and $T_g$ follows a power-law behavior
    with exponent $1.5(1)$.  { Different system sizes are
      given by different symbols in which filled and hollow symbols
      refer to $T_g$ and $T_e$, respectively. The color code
      corresponds to different shear rates $\dot\gamma = 0.02$
      (purple), $0.04$ (magenta), $0.06$ (blue), $0.08$ (golden), and
      $0.1$ (yellow).} ({\bf b}) The collapse of all data presented in
    Fig.~\ref{fig:FT}-a when the vertical axis is multiplied by a
    factor of $T_g/\tau$ with $\tau=0.28$. { In these
      data, we cover a large range of packing fractions around
      jamming, $0.7<\phi<0.9$.} { Error bars correspond to
      square root of variance.}
    \label{fig:T_eff_master}}
\end{figure}

\end{document}